\documentclass{emulateapj}

\usepackage{lscape}
\usepackage{apjfonts}
\usepackage{graphicx}
\usepackage{natbib}
\newcommand{\etal}{et al.}
\newcommand{\hbeta}{H{$\beta$}}
\newcommand{\halpha}{H{$\alpha$}}
\newcommand{\CIV}{C{\sevenrm IV}}

\newcommand{\MgII}{Mg{\sevenrm II}}

 \font\sevenrm=cmr7 scaled 1000

\begin{document}

\title{The Impact of the Uncertainty in Single-Epoch Virial Black Hole Mass Estimates on the
Observed Evolution of the Black Hole -- Bulge Scaling Relations}

\shorttitle{THE VIRIAL MASS BIAS}

\slugcomment{Draft Version}

\shortauthors{SHEN AND KELLY}
\author{Yue Shen and Brandon C. Kelly\altaffilmark{1}\\
Harvard-Smithsonian Center for Astrophysics, 60 Garden St., MS-51,
Cambridge, MA 02138, USA} \altaffiltext{1}{Hubble Fellow}

\begin{abstract}
Recent observations of the black hole (BH) - bulge scaling
relations usually report positive redshift evolution, with higher
redshift galaxies harboring more massive BHs than expected from
the local relations. All of these studies focus on broad line
quasars with BH mass estimated from virial estimators based on
single-epoch spectra. Since the sample selection is largely based
on quasar luminosity, the cosmic scatter in the BH-bulge relation
introduces a statistical bias leading to on average more massive
BHs given galaxy properties at high redshift
\citep{Lauer_etal_2007b}. We here emphasize a previously
under-appreciated statistical bias resulting from the uncertainty
of single-epoch virial BH mass estimators and the shape of the
underlying (true) BH mass function, which leads to on average
overestimation of the true BH masses at the high-mass end
\citep{Shen_etal_2008b}. We demonstrate that the latter virial
mass bias can contribute a substantial amount to the observed
excess in BH mass at fixed bulge properties, comparable to the
Lauer \etal\ bias. The virial mass bias is independent of the
Lauer et al.\ bias, hence if both biases are at work, they can
largely (or even fully) account for the observed BH mass excess at
high redshift.
\end{abstract}

\section{Introduction}\label{sec:intro}

The redshift evolution of the local scaling relations between
galaxy bulge properties and the mass of the central supermassive
black hole (SMBH) has important clues to the establishment of
these tight relations across cosmic time. There is currently a
huge effort in measuring this evolution, either in terms of the BH
mass--bulge velocity dispersion ($M_\bullet-\sigma$) relation, or
the BH mass--bulge stellar mass/luminosity ($M_\bullet-M_{\rm
bulge}$, $M_\bullet-L_{\rm bulge}$) relation
\citep[e.g.,][]{Shields_etal_2003,Peng_etal_2006a,Peng_etal_2006b,Salviander_etal_2007,ShenJ_etal_2008,
Woo_etal_2006, Woo_etal_2008, Treu_etal_2007,Jahnke_etal_2009,
Greene_etal_2009,
Merloni_etal_2009,Decarli_etal_2009b,Bennert_etal_2009}. With a
few exceptions, most of these studies report a strong positive
evolution in BH mass for fixed bulge properties, which reaches as
high as $\sim 0.6$ dex offset in BH mass from the local relation
at the highest redshift probed ($z\sim 2$).

This observed strong evolution is difficult to understand in two
aspects. For the $M_\bullet-\sigma$ relation, numerical
simulations based on self-regulated BH growth
\citep[e.g.,][]{Robertson_etal_2006} and other theoretical
arguments \citep[e.g.,][]{Shankar_etal_2009c} favor a mild
evolution in the normalization of the $M_\bullet-\sigma$ relation,
in conflict with the otherwise claimed strong evolution. For the
$M_\bullet-M_{\rm bulge}$ or $M_\bullet-L_{\rm bulge}$ relation,
many of the observed hosts at earlier epoches are already bulge-dominated\footnote{We note that there are also some observed hosts which are likely to be late-type \citep[e.g.,][]{Sanchez_etal_2004,Letawe_etal_2007,Merloni_etal_2009}. },
hence unless their bulges continue to grow a considerable amount over time, it is difficult to understand how these systems would have migrated to the local relation.

An alternative explanation for this observed strong evolution is
that it is caused, at least partly, by the systematics involved in
deriving both host properties and BH masses. All these studies
mentioned above focus on broad line quasar samples, with host
properties measured from the quasar-light subtracted images or
spectra. The systematics in subtracting the quasar light, as well
as in converting observables (such as photometric colors or
spectral properties) to physical quantities (such as bulge stellar mass) is a
potential contamination to the final results. More importantly,
the samples are predominately selected in quasar luminosity and
hence are {\em not} unbiased samples for studies of the BH scaling
relations.

One statistical bias resulting from using quasar-luminosity
selected samples was discussed in detail in
\citet{Lauer_etal_2007b}. Because there is cosmic scatter in the
BH scaling relations and because the galaxy luminosity (or stellar
mass, velocity dispersion, etc) function is bottom heavy, there
are more smaller galaxies hosting the same mass BHs than the more
massive galaxies. Hence when selecting samples based on quasar
luminosities, low-mass BHs are under-represented, leading to an
apparent excess in BH mass at fixed host properties. If the cosmic
scatter in BH scaling relations increases with redshift and
reaches $\ga 0.5$ dex, it can in principle account for all the
excess in BH mass observed for the most luminous quasars
\citep[e.g.,][]{Lauer_etal_2007b,Merloni_etal_2009}.

A second statistical bias, resulting from the uncertainty in the
BH mass estimates, was pointed out in Shen \etal\ (2008b, also see
Kelly \etal\ 2009a). In all these studies, the BH masses are
estimated using the so-called virial method based on single-epoch
spectra. In this method, one assumes that the broad line region
(BLR) is virialized, and the BH mass can be estimated as $M_{\rm
vir}\approx G^{-1}RV^2$, where $R$ is the BLR radius and $V$ is
the virial velocity; one further estimates the BLR radius using a
correlation between $R$ and the continuum luminosity $L$, i.e.,
$R\propto L^{C_1}$, found in local reverberation mapping (RM)
samples
\citep[e.g.,][]{Kaspi_etal_2005,Bentz_etal_2006,Bentz_etal_2009},
and estimates the virial velocity using the width of the broad
lines. In this way, one can estimate a virial BH mass using
single-epoch spectra: $\log M_{\rm vir}= \log R + 2\log({\rm
FWHM})+ {\rm const}= C_1\log L+2\log({\rm FWHM})+{\rm const}$. The
coefficients are calibrated empirically using RM samples or
inter-calibrations between various lines
\citep[e.g.,][]{McLure_Jarvis_2002,Vestergaard_Peterson_2006,Mcgill_etal_2008}.

Even though the virial method is widely used, these BH mass
estimates based on several lines (usually \halpha, \hbeta, \MgII\
and \CIV) have a non-negligible uncertainty $\sigma_{\rm vir}\ga
0.4$ dex, when compared to RM masses or masses derived from the
$M_\bullet-\sigma$ relation
\citep[e.g.,][]{Vestergaard_Peterson_2006,Onken_etal_2004}. This
uncertainty must come from the imperfectness of using luminosity
and line width as proxies for the BLR radius and virial velocity,
i.e., there are substantial {\em uncorrelated} variations
$\sigma_{L}$ in luminosity and variations $\sigma_{\rm FWHM}$ in
line width, which together contribute to the overall uncertainty
$\sigma_{\rm vir}$, i.e.,
\begin{equation}\label{eqn:sig_vir}
\sigma_{\rm vir}=\sqrt{(C_1\sigma_{L})^2 + (2\sigma_{\rm
FWHM})^2}\ .
\end{equation}
One can then imagine a statistical bias will arise if the
underlying active black hole mass function (BHMF) is bottom heavy.
In particular, the variations (uncompensated by the variations in
FWHM) of luminosity at fixed true BH mass, $\sigma_L$, will
scatter more lighter BHs into a luminosity bin than heavier BHs,
and bias the mean BH mass in that bin. This is the Malmquist-type
bias (or Eddington bias) emphasized in \citet[][sec
4.4]{Shen_etal_2008b}, which has received little attention in the
studies on the evolution of BH scaling relations to date.

In this paper we examine the impact of this mass bias on the
observed evolution in BH scaling relations with realistic models
for the underlying true BHMF and quasar luminosity function (LF).
In \S\ref{sec:mass_bias} we review the statistical mass bias
discussed in \citet[][]{Shen_etal_2008b} and demonstrate its
effects with simple models; in \S\ref{sec:sim} we consider more
realistic intrinsic BHMF and quasar LF, and estimate the mass bias
as functions of luminosity and redshift; we discuss the impact of
this mass bias and conclude in \S\ref{sec:disc}. Throughout the
paper we use cosmological parameters $\Omega_0=0.3$,
$\Omega_\Lambda=0.7$ and $H_0=70\ {\rm km\,s^{-1}\,{Mpc}^{-1}}$.
Luminosities are in units of ${\rm erg\,s^{-1}}$ and BH masses are
in units of $M_\odot$. Unless otherwise specified, ``luminosity''
refers to the bolometric luminosity, and we are only concerned
with the active BH population.

\section{The Black Hole Mass Bias}\label{sec:mass_bias}
Denoting $\lambda\equiv\log L$ where $L$ is quasar luminosity
(bolometric or in a specific band), $m\equiv\log M_\bullet$ the
true BH mass, and $m_e\equiv\log M_{\rm vir}$ the virial BH mass,
the probability distribution of virial mass $m_e$ given true mass
$m$ is:
\begin{equation}\label{eqn:mvir_mtrue}
p_0(m_e|m)=(2\pi\sigma_{\rm
vir}^2)^{-1/2}\exp\bigg[-\frac{(m_e-m)^2}{2\sigma_{\rm
vir}^2}\bigg]\ ,
\end{equation}
where $\sigma_{\rm vir}$ is the uncertainty of the virial BH mass.

The probability distribution of true BH mass $m$ given virial mass
$m_e$ is:
\begin{equation}\label{eqn:mtrue_mvir}
p_1(m|m_e)=p_0(m_e|m)\Psi_M(m)\bigg(\int
p_0(m_e|m)\Psi_M(m)dm\bigg)^{-1}\ ,
\end{equation}
where $\Psi_M(m)\equiv dn/dm$ is the true BHMF. If the sample is
selected irrespective of luminosity (i.e., no flux limit) and so
we can see all BHs with arbitrary $m$, then for a power-law true
BHMF $\Psi_M(m)\propto 10^{m\gamma_M}$ and at fixed $m_e$, the
statistical bias between the expectation value $\langle m\rangle$
and virial mass $m_e$ is simply:
\begin{equation}\label{eqn:bias_M}
\Delta\log M_{\bullet}= m_e - \langle m \rangle
=-\ln(10)\gamma_{M}\sigma_{\rm vir}^2\ .
\end{equation}

In reality the sample is usually selected in quasar luminosity. If
the distribution of luminosity at a given true BH mass for broad
line quasars is $g(\lambda|m)$ where $\int
g(\lambda|m)d\lambda=1$, the expectation value of $m$ at fixed
luminosity $\lambda$ is:
\begin{equation}\label{eqn:true_avg_fix_L}
\langle m\rangle_{\lambda}=\frac{\int
mg(\lambda|m)\Psi_M(m)\,dm}{\int g(\lambda|m)\Psi_M(m)\,dm}\ ,
\end{equation}
and the luminosity weighted average of $m$ in the quasar sample
is:
\begin{equation}\label{eqn:true_avg}
\langle m\rangle=
\frac{\int_{\lambda_1}^{\lambda_2}\Psi_L(\lambda)\ \langle
m\rangle_{\lambda}\,d\lambda}{\int_{\lambda_1}^{\lambda_2}\Psi_L(\lambda)\,d\lambda}\
,
\end{equation}
where $\lambda_1$ and $\lambda_2$ are the luminosity limits in the
sample, and the quasar luminosity function $\Psi_L\equiv
dn/d\lambda$ is
\begin{equation}\label{eqn:LF}
\Psi_{L}(\lambda)=\int\ \Psi_{M}(m)g(\lambda|m)\,dm\ .
\end{equation}

On the other hand, the virial BH masses depend on luminosities,
and the sample averaged virial mass is:
\begin{equation}\label{eqn:vir_avg}
\langle
m_e\rangle=\frac{\int_{\lambda_1}^{\lambda_2}\Psi_L(\lambda)\
\frac{\int m_ef(m_e|\lambda)\,dm_e}{\int
f(m_e|\lambda)\,dm_e}\,d\lambda}{\int_{\lambda_1}^{\lambda_2}\Psi_L(\lambda)\,d\lambda}\
,
\end{equation}
where $f(m_e|\lambda)$ is the probability distribution of $m_e$
given luminosity $\lambda$ and $\int f(m_e|\lambda)\,dm_e=1$. In
the absence of a theoretical model for $f(m_e|\lambda)$, here we
adopt an empirical recipe for $f(m_e|\lambda)$: recall that the
virial BH mass is expressed in terms of luminosity and FWHM.
Neglecting the scatter from converting bolometric luminosity to
continuum luminosity, which is typically $\sim 0.1$ dex
\citep[e.g.,][]{Richards_etal_2006b}, we have
\begin{equation}
m_e=C_1\lambda + 2\log({\rm FWHM}) + C_2\ .
\end{equation}
Here $C_1\approx 0.5-0.7$ is the slope in the measured
luminosity-radius relation for BLRs and $C_2$ is calibrated
empirically
\citep[e.g.,][]{Kaspi_etal_2005,Kaspi_etal_2007,Bentz_etal_2006,Bentz_etal_2009,McLure_Jarvis_2002,Vestergaard_Peterson_2006}.
For statistical quasar samples, the distribution of FWHMs does not
depend on luminosity much 
\citep[e.g.,][]{Shen_etal_2008b,Fine_etal_2008}, i.e., for any
given luminosity, the FWHM distribution has a constant mean value
and a log-normal scatter $\sigma_{\rm FWHM}$ around the mean.
Therefore at fixed luminosity, the value of $m_e$ is independent
of the true mass $m$, i.e.,
\begin{equation}\label{eqn:me_lambda}
f(m_e|\lambda)=(2\pi \sigma_{\rm
line}^2)^{-1/2}\exp\bigg\{-\frac{[m_e-(C_1\lambda+C_2)]^2}{2\sigma_{\rm
line}^2}\bigg\}\ ,
\end{equation}
where $\sigma_{\rm line}=2\sigma_{\rm FWHM}$ is the scatter in
virial mass resulting from the scatter in FWHM ($\sigma_{\rm
FWHM}$) at fixed luminosity, and the constant mean value of FWHM
is absorbed in $C_2$.


Eqns.\ (\ref{eqn:true_avg})-(\ref{eqn:vir_avg}) have analytical
results for specific forms of $\Psi_M(m)$ and $g(\lambda|m)$. For
a power-law true BH mass function $\Psi_M(m)\propto
10^{m\gamma_M}$, and a power-law relation between true mass and
luminosity $\langle\lambda\rangle=am+b$ with a constant log-normal
scatter $\sigma_L$, i.e.,
\begin{equation}\label{eqn:g1}
g(\lambda|m)=(2\pi
\sigma_L^2)^{-1/2}\exp\bigg\{-\frac{[\lambda-(am+b)]^2}{2\sigma_L^2}\bigg\}\
,
\end{equation}
Eqns.\ (\ref{eqn:true_avg})-(\ref{eqn:vir_avg}) yield
\begin{equation}
\Psi_L(\lambda)\propto 10^{\lambda\gamma_M/a}\ ;
\end{equation}
\begin{eqnarray}\label{avg_m_pl}
\langle
m\rangle=\frac{\gamma_M\ln(10)\sigma_L^2-ab}{a^2}&-&\frac{1}{\gamma_M\ln(10)}\nonumber\\
&+&\frac{\lambda_2 10^{\lambda_2\gamma_M/a}-\lambda_1
10^{\lambda_1\gamma_M/a}}{a(10^{\lambda_2\gamma_M/a}-10^{\lambda_1\gamma_M/a})}\
;
\end{eqnarray}
\begin{eqnarray}\label{avg_me_pl}
\langle m_e\rangle=C_2&-&\frac{C_1a}{\gamma_M\ln(10)}\nonumber\\
&+&\frac{C_1(\lambda_2 10^{\lambda_2\gamma_M/a}-\lambda_1
10^{\lambda_1\gamma_M/a})}{10^{\lambda_2\gamma_M/a}-10^{\lambda_1\gamma_M/a}}\
.
\end{eqnarray}
Note throughout this paper we assume this constant scatter in
luminosity, $\sigma_L$, is solely the variations that are
uncompensated by the corresponding variations in FWHM, i.e., it is
one of the two sources of the virial uncertainty $\sigma_{\rm
vir}$ (Eqn.\ \ref{eqn:sig_vir})\footnote{It is possible to
incorporate an additional {\em correlated} variation term in
luminosity at fixed true mass, which is compensated by the
variations in FWHM and hence does not add to the virial
uncertainty, but this will makes the mass bias discussed in
\S\ref{sec:sim} even more severe (see later discussions).}.

There are eight parameters that determine the bias $\Delta\log
M_{\bullet}=\langle m_e \rangle - \langle m\rangle$: $\sigma_L$,
$\gamma_M$, $a$ and $b$ are for the underlying true BH mass
function and luminosity (or Eddington ratio) distributions at
fixed true mass, which are model assumptions and must be tuned to
produce the observed quasar luminosity function; $C_1$ and $C_2$
are determined from the empirically calibrated virial estimators
and measured FWHM distributions in statistical quasar samples;
$\lambda_1$ and $\lambda_2$ are observational windows determined
from the specific quasar sample under investigation. A further
constraint comes from the assumption that virial mass is an unbiased
estimator of the true BH mass (e.g., Eqn.\ \ref{eqn:mvir_mtrue}),
which requires (e.g., consulting Eqns.\ \ref{eqn:mvir_mtrue},
\ref{eqn:me_lambda} and \ref{eqn:g1})
\begin{equation}\label{eqn:ab_C}
C_1=1/a,\qquad C_2=-b/a\ .
\end{equation}
Thus the bias reduces to
\begin{equation}\label{eqn:mass_bias}
\Delta\log M_{\bullet}=-\gamma_M\ln(10)\sigma_L^2/a^2\ ,
\end{equation}
from Eqns.\ (\ref{avg_m_pl}) and (\ref{avg_me_pl}) \citep[see also
sec 4.4 in][]{Shen_etal_2008b}, which is independent on
luminosity.

For demonstration purposes here we take a model with parameters
$\sigma_L=0.4$, $\gamma_M=-4$, $a=2$ and $b=28.4$. These
parameters produce a slope of $\gamma_L=-2$ in the quasar LF,
close to the observed bright-end slope
\citep[e.g.,][]{Richards_etal_2006a,Hopkins_etal_2007a}, and
constants $C_1=1/a=0.5$ and $C_2=-b/a=-14.2$, which are consistent
with the commonly used virial mass calibrations and the observed
FWHM distributions
\citep[e.g.,][]{McLure_Jarvis_2002,Vestergaard_Peterson_2006,Mcgill_etal_2008,Shen_etal_2008b}.
With these parameters, the difference between the sample-averaged
true and virial BH masses is $\langle m_e \rangle - \langle
m\rangle\approx 0.37$ dex.

The derivation of this virial mass bias is independent from the
derivation of the bias discussed in \citet{Lauer_etal_2007b}. To
illustrate this, let us consider again simple power-law
distributions and Gaussian (lognormal) scatter. We start from a
power-law distribution of quasar host galaxy property $s$
($s\equiv \log M_{\rm bulge}$ for instance),
\begin{equation}\label{eqn:p_s}
\Psi_S(s)\propto 10^{s\gamma_S}\ .
\end{equation}
The distribution of true BH mass $m$ at fixed $s$ is:
\begin{equation}\label{eqn:p_m_s}
p_0(m|s)=(2\pi\sigma_{\mu}^2)^{-1/2}\exp\bigg\{-\frac{[m-(C_3s+C_4)]^2}{2\sigma_{\mu}^2}\bigg\}\
,
\end{equation}
where the BH scaling relation is $\langle m\rangle=C_3s+C_4$ with
a cosmic scatter $\sigma_{\mu}$, where $C_3$ and $C_4$ are
constants. Given the power-law distribution of $s$ (Eqn.\
\ref{eqn:p_s}), the distribution of true BH mass $m$ is then also
a power-law $\Psi_M(m)\propto 10^{m\gamma_M}$ where
$\gamma_M=\gamma_S/C_3$. Assuming the distribution of luminosity
$\lambda$ at fixed $m$ is given by Eqn.\ (\ref{eqn:g1}) we can
derive the distribution of $\lambda$ at fixed $s$ as:
\begin{eqnarray}\label{eqn:p_lambda_s}
p_0(\lambda|s)&=&\int g(\lambda|m)p_0(m|s) dm\nonumber\\
&=&[2\pi(\sigma_L^2+a^2\sigma_{\mu}^2)]^{-1/2}\exp\bigg\{-\frac{\displaystyle[\lambda-(aC_3s+b+aC_4)]^2}{\displaystyle
2(\sigma_L^2+a^2\sigma_{\mu}^2)}\bigg\}\ . \nonumber\\
\end{eqnarray}
Therefore the distribution of $s$ at fixed luminosity $\lambda$ is
\begin{eqnarray}\label{eqn:p_s_lambda}
p_1(s|\lambda)&=&p_0(\lambda|s)\Psi_S(s)\bigg(\int
p_0(\lambda|s)\Psi_S(s)ds\bigg)^{-1}\nonumber\\
&=&(2\pi\sigma_1^2)^{-1/2}\exp\bigg\{-\frac{[s-(C_5\lambda+C_6)]^2}{2\sigma_1^2}\bigg\}
,
\end{eqnarray}
where
\begin{eqnarray}
\sigma_1^2=\frac{\sigma_L^2+a^2\sigma_\mu^2}{a^2C_3^2},\
C_5=\frac{1}{aC_3},\ C_6=\ln(10)\gamma_S\sigma_1^2 -
\frac{b+aC_4}{aC_3}\ .\nonumber\\
\end{eqnarray}
Similarly the distribution of $m$ at fixed $\lambda$ is:
\begin{eqnarray}\label{eqn:p_m_lambda}
p_1(m|\lambda)&=&g(\lambda|m)\Psi_M(m)\bigg(\int
g(\lambda|m)\Psi_M(m)dm\bigg)^{-1}\nonumber\\
&=&(2\pi\sigma_2^2)^{-1/2}\exp\bigg\{-\frac{[m-(C_7\lambda+C_8)]^2}{2\sigma_2^2}\bigg\}\
,
\end{eqnarray}
where
\begin{eqnarray}
\sigma_2^2=\frac{\sigma_L^2}{a^2},\ C_7=\frac{1}{a},\
C_8=\frac{\ln(10)\gamma_S\sigma_L^2}{a^2C_3}-\frac{b}{a}\ .
\end{eqnarray}
Hence the Lauer \etal\ bias, i.e., the excess in true BH mass at
fixed luminosity is simply:
\begin{eqnarray}\label{eqn:lauer_bias}
\Delta\log M_{\bullet,{\rm Lauer}}&=&\langle m \rangle|_\lambda -
(C_3\langle s \rangle|_\lambda + C_4)\nonumber\\
&=&-\ln(10)\gamma_S\sigma_\mu^2/C_3\ .
\end{eqnarray}
Since $\gamma_S$ is negative, the Lauer \etal\ bias states that at
fixed luminosity, the average true BH mass is already biased high
with respect to the expectation from the mean BH scaling relation.
On top of that, the virial mass bias (Eqn.\ \ref{eqn:mass_bias})
states that the average BH mass estimate is further biased high
with respect to the true mass.

\begin{figure}
  \centering
    \includegraphics[width=0.48\textwidth]{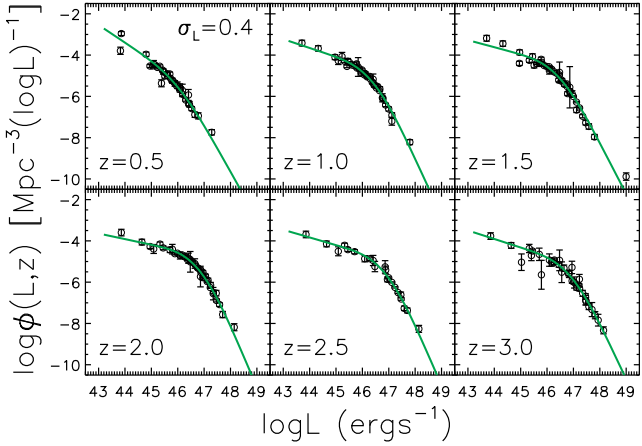}
    \caption{Model quasar LF for a nominal value of $\sigma_L=0.4$ dex and at several redshifts (green lines).
    The points are the derived bolometric LF data based on optical and soft X-ray data compiled by \citet[][and references therein]{Hopkins_etal_2007a}.}
    \label{fig:LF}
\end{figure}

\begin{figure}
  \centering
    \includegraphics[width=0.48\textwidth]{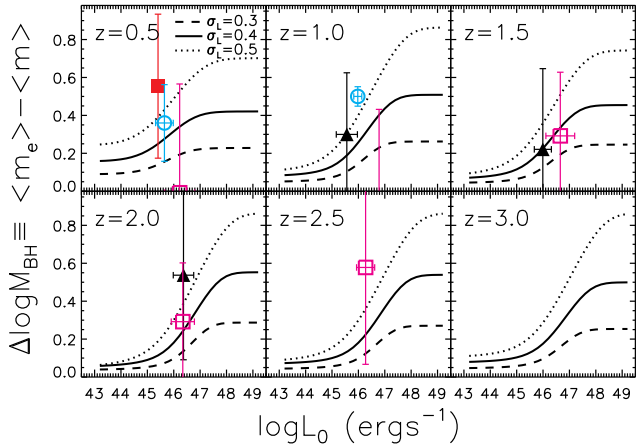}
    \caption{Estimated BH mass bias using our best-fit models of the true BHMF and
    luminosity distributions, for luminosity-limited quasar samples with $\lambda>\lambda_0\equiv \log L_0$ at various redshifts.
    Shown here are the results for three fixed values of $\sigma_L=0.3$, 0.4 and 0.5 dex. The bias increases
    when the luminosity or $\sigma_L$ increases. We also plot the observed BH mass excess at fixed
    host properties from \citet[][filled square]{Woo_etal_2006,Woo_etal_2008}, \citet[][open circles]{Salviander_etal_2007},
    \citet[][filled trianges]{Merloni_etal_2009}, and \citet[][open squares]{Decarli_etal_2009b} at the corresponding
    redshifts. Error bars are the
    standard deviations of objects in their samples. Note that the data from \citet{Salviander_etal_2007} are already
    binned, hence the dispersion is substantially smaller.}
    \label{fig:bias}
\end{figure}

\section{More Realistic Models}\label{sec:sim}
The simple estimate of the BH mass bias discussed above neglects
the curvature in the intrinsic BHMF and quasar LF, hence is only
valid at the bright end of the LF. In particular since the quasar
LF flattens below the break luminosity $\lambda_{\rm break}\sim
46$, we expect that the mass bias becomes less severe towards
fainter luminosities. It is important to realize that in
essentially all of the studies mentioned in \S\ref{sec:intro},
larger offsets usually occur at higher redshift, when their
samples are sampling the higher-mass end of the intrinsic BHMF. It
is conceivable then that a false evolution will arise simply from
the increasing mass bias towards the high-mass tail.

To estimate the BH mass bias at various redshifts and luminosity
sampling ranges, we assume an underlying BHMF and a model for the
luminosity distribution at fixed true BH mass\footnote{We note
that in principle one should use an underlying BHMF that is
constrained by the measured BHMF based on virial masses
\citep[see, e.g.,][]{Kelly_etal_2009a,Kelly_etal_2009b}.
Unfortunately, the derived BHMF is poorly constrained at
$M_\bullet\la 10^8\ M_\odot$ for current flux-limited quasar samples.},
using the observed quasar LF as constraints. This forward-modeling
procedure is similar to the simulations done in
\citet[][]{Shen_etal_2008b} and
\citet{Kelly_etal_2009a,Kelly_etal_2009b}.

We start with a broken power-law model for the true BHMF:
\begin{equation}
\Psi_{M}(m)\propto
\frac{1}{10^{-\gamma_{M1}(m-m_*)}+10^{-\gamma_{M2}(m-m_*)}}\ ,
\end{equation}
where $\gamma_{M1}$ and $\gamma_{M2}$ are the low-mass and
high-mass end slope, and $m_*$ is the break BH mass. For the
luminosity distribution $g(\lambda|m)$ we still use the Gaussian
form in Eqn.\ (\ref{eqn:g1}). For simplicity, we assume that the
set of six parameters, $\gamma_{M1}$, $\gamma_{M2}$, $m_*$, $a$,
$b$ and $\sigma_L$, are only functions of redshift. At a fixed
redshift, a set of the six parameters determines the shape of the
quasar LF (\ref{eqn:LF}), which is to be constrained by
observations. Assuming that the commonly used virial BH estimators
are unbiased, we fix $a=2$ and $b=28.4$, which corresponds to a
BLR radius-luminosity relation $R\propto L^{0.5}$ and virial
coefficients and FWHM distributions consistent with previous work
\citep[e.g.,][]{Shen_etal_2008b,Fine_etal_2008}. We also fix the
value of $\sigma_L$ during model fitting. Therefore there are four
free parameters: $\gamma_{M1}$, $\gamma_{M2}$, $m_*$, and the
normalization of the BHMF, $\Phi_0$ (in units of ${\rm Mpc}^{-3}{\rm
dex}^{-1}$). We list the best-fit model parameters in Table 1 for different
values of $\sigma_L$ and at several redshifts. We note that these models
are mainly used to demonstrate the effects of the virial mass bias, and
should not be interpreted as our attempt to constrain the true active BHMF.

\begin{deluxetable}{rrrrrr}
\tablecolumns{6} \tablewidth{0.45\textwidth}
\tablecaption{Model Parameters
\label{table:notation}} \tablehead{$\sigma_{L}$ & $z$ & $\gamma_{M1}$ & $\gamma_{M2}$ & $m_{*}$ & $\log\Phi_0$}
\startdata
0.3 & 0.5 & -1.72 & -4.40 & 8.61 & -4.46\\
      & 1.0 & -0.95 & -5.06 & 8.81 & -4.19\\
      & 1.5 & -0.83 & -4.74 & 8.88 & -4.06\\
      & 2.0 & -0.71 & -5.54 & 9.03 & -4.30\\
      & 2.5 & -0.82 & -5.22 & 9.04 & -4.42\\
      & 3.0 & -0.88 & -4.89 & 9.06 & -4.59\\
\hline\\
0.4 & 0.5 & -1.70 & -4.57 & 8.52 & -4.45\\
      & 1.0 & -0.85 & -5.52 & 8.72 & -4.18\\
      & 1.5 & -0.69 & -4.93 & 8.77 & -4.01\\
      & 2.0 & -0.54 & -6.00 & 8.93 & -4.24\\
      & 2.5 & -0.74 & -5.85 & 8.96 & -4.42\\
      & 3.0 & -0.83 & -5.42 & 8.98 & -4.62\\
\hline\\
0.5 & 0.5 & -1.68 & -4.88 & 8.40 & -4.39\\
      & 1.0 & -0.68 & -6.00 & 8.59 & -4.05\\
      & 1.5 & -0.50 & -5.16 & 8.63 & -3.87\\
      & 2.0 & -0.23 & -6.00 & 8.77 & -4.05\\
      & 2.5 & -0.52 & -6.00 & 8.80 & -4.25\\
      & 3.0 & -0.71 & -6.00 & 8.86 & -4.52\\
\enddata
\tablecomments{The $\chi^2$ fits are done at individual redshifts against
the bolometric LF data compiled in \citet{Hopkins_etal_2007a}, i.e., we do not fit a global broken power-law BHMF model. For certain values of $\sigma_L$ and
redshifts, the high-mass end slope is poorly constrained, in which case we fixed
the high-mass end slope to be $\gamma_{M2}=-6$.}
\end{deluxetable}

Fig.\ \ref{fig:LF} shows the model LF with fixed $\sigma_L=0.4$
and comparison with the bolometric LF data compiled in
\citet[][and references therein]{Hopkins_etal_2007a}, at several
redshifts. At each redshift, we determine $\gamma_{M1}$,
$\gamma_{M2}$, $m_*$ and $\Phi_0$ by minimizing the $\chi^2$
between the model and the data. As discussed earlier, this scatter
$\sigma_L$ refers to the variations that are uncompensated by the
corresponding variations in FWHM. If we were to incorporate an
additional {\em correlated} variation term in luminosity at fixed
true mass, which are compensated by an additional variation in
FWHM, we would need a steeper high-mass end slope for the true
BHMF in order not to overshoot the bright-end LF, which then makes
the mass bias even worse. Also, the amount of this additional
broadening is limited by the already narrow distributions of FWHM
\citep[e.g.,][]{Shen_etal_2008b,Fine_etal_2008} and the
requirement that $\sigma_{\rm vir}\ga 0.4$.

Given the model true BHMF and luminosity distribution, we can now
estimate the bias $\Delta\log M_{\bullet}=\langle m_e \rangle -
\langle m\rangle$ as a function of the sampled luminosity range
using Eqns.\ (\ref{eqn:true_avg}) and (\ref{eqn:vir_avg}). Fig.\
\ref{fig:bias} shows $\Delta\log M_{\bullet}$ for
luminosity-limited quasar samples with
$\lambda>\lambda_0\equiv\log L_0$ at several redshifts and for
fixed $\sigma_L=0.3$, 0.4, 0.5 dex. The bias generally rises when
luminosity increases and approaches the asymptotic value $-\gamma_{M2}\ln(10)\sigma_L^2/a^2$ at the bright end. Because we
are fitting our model against the bolometric LF data at individual
redshifts, we do not expect identical results since there will be
systematics involved in deriving the bolometric LF data from band
conversions and from the assumed quasar spectral energy
distribution \citep[see details in ][]{Hopkins_etal_2007a}.
Nevertheless, this analysis demonstrates that the mass bias
$\langle m_e\rangle-\langle m\rangle$ could be substantial for
bright luminosities and for large dispersion $\sigma_L$ in the
luminosity distribution at fixed true mass.

\section{Discussion}\label{sec:disc}

The virial mass bias discussed here depends on the amplitude of
the scatter $\sigma_L$. This is the variation in luminosity at
fixed true BH mass which is {\em not compensated} by the variation
in FWHM. How large is $\sigma_L$? For the best studied local RM
samples and for \hbeta\ only
\citep[e.g.,][]{Kaspi_etal_2005,Bentz_etal_2006,Bentz_etal_2009},
the scatter of luminosity around fixed BLR size is on the order of
$\la 0.2$ dex. However, this is for the best cases. We generally
expect larger $\sigma_L$ for statistical quasar samples with
single-epoch spectra for the following reasons: 1) single-epoch
spectra have larger scatter than averaged spectra in RM samples;
2) these samples usually cover a luminosity range that is poorly
sampled by current RM data; 3) the UV regime of the spectrum (in
particular for \CIV) has more peculiarities and larger scatter in
the $R-L$ relation than the \hbeta\ regime, but is usually used at
high redshift due to the drop off of \hbeta\ in the optical
spectrum; 4) even if some variations in luminosity truthfully
trace the variations in BLR radius $R$, they may still be
uncompensated by FWHM if, for instance, there is a non-virial
component in the broad line which does not respond to BLR radius
variations \citep[especially for \CIV, see discussions in
][]{Shen_etal_2008b}; 5) in many statistical quasar samples, the
spectra have low signal-to-noise ratios, which lead to increased
uncorrelated scatter between the measured luminosity and FWHM, and
bias the virial mass estimates more. Hence it is very plausible
that $\sigma_L\ga 0.4$ in most cases, which then accounts for $\ga
0.2$ dex in $\sigma_{\rm vir}$ (e.g., Eqn.\ \ref{eqn:sig_vir}).

Therefore we conclude that the mass bias discussed here
contributes at least $0.2-0.3$ dex at $L_{\rm bol}\ga 10^{46}\
{\rm erg\,s^{-1}}$, comparable to the statistical bias resulting
from the cosmic scatter in the BH scaling relations
\citep[e.g.,][]{Lauer_etal_2007b}. In Fig.\ \ref{fig:bias} we also
show the ``observed'' BH mass offset (excess at fixed galaxy
properties) from the literature. We note that the amount of the
uncorrelated scatter $\sigma_L$ might increase with redshift both
due to the switch from the \hbeta\ line to the more problematic UV
lines and due to the often decreased spectral quality at high
redshift. Since the virial mass bias is independent of the Lauer
\etal\ bias, the two statistical biases together can account for a
large (or even full) portion of the BH mass excess seen in the
data, hence no exotic scenarios are needed to explain this strong
redshift evolution. It is possible, however, that there is still a
mild evolution in these BH scaling relations, which is inherent to
the cosmic co-formation of SMBHs and their hosts. But given these
statistical biases, and given other systematics with single-epoch
virial estimators \citep[not discussed here, but see,
e.g.,][]{Denney_etal_2009}, it is premature to claim a strong
positive evolution in the BH scaling relations.

\acknowledgements

We thank the anonymous referee, as well as Tod Lauer, Scott Tremaine, Jenny Greene, and Avi Loeb for useful comments. YS acknowledges support from a Clay Postdoctoral Fellowship through the Smithsonian Astrophysical Observatory
(SAO). BK acknowledges support by NASA through Hubble Fellowship
grant \#HF-51243.01 awarded by the Space Telescope Science
Institute, which is operated by the Association of Universities
for Research in Astronomy, Inc., for NASA, under contract NAS
5-26555.


\end{document}